\documentclass[a4paper,11pt]{article}
\usepackage{pos}

\newcommand{\gtapprox}{\raisebox{-0.5ex}{$\,\stackrel{>}{\scriptstyle\sim}\,$}}

\title{Computation of masses of quarkonium bound states using heavy quark potentials from lattice QCD}
\ShortTitle{Computation of masses of quarkonium bound states}

\author*[a,b]{Michael Eichberg}
\author[a,b]{Marc Wagner}

\affiliation[a]{Institut f\"ur Theoretische Physik, Goethe-Universit\"at Frankfurt am Main,\\
  Max-von-Laue-Straße 1, D-60438
  Frankfurt am Main, Germany}

\affiliation[b]{Helmholtz Research Academy Hesse for FAIR, Campus Riedberg, \\
	 Max-von-Laue-Straße 12, D-60438 
	 Frankfurt am Main, Germany}

\emailAdd{eichberg@itp.uni-frankfurt.de}
\emailAdd{mwagner@itp.uni-frankfurt.de}

\abstract{
	We compute masses of bottomonium and charmonium bound states using a Schr\"odinger equation with a heavy quark-antiquark potential including $1/m$ and $1/m^2$ corrections previously derived in potential Non-Relativistic QCD and computed with lattice QCD. This is a preparatory step for a future project, where we plan to take into account similar corrections to study quarkonium resonances and tetraquarks above the lowest meson-meson thresholds.
    }

\FullConference{%
  FAIR next generation scientists - 7th Edition Workshop (FAIRness2022)\\
  23-27 May 2022\\
  Paralia (Pieria, Greece)
}

\begin{document}
\maketitle

\section{Introduction}

    Heavy quarkonium systems are characterized by small relative velocities between the heavy quark and the heavy antiquark. Such systems are suited for investigation within the framework of potential Non-Relativistic QCD (pNRQCD), which is an effective field theory based on a multipole expansion of the full QCD Lagrangian in the relative quark velocity $v_Q$. As such, it describes $b\bar{b}$ systems ($v_b^2 \sim 0.1$) rather precisely and $c\bar{c}$ systems ($v_c^2\sim 0.3$) still in a reasonable approximation. The pNRQCD Lagrangian contains potentials, which can be expressed in terms of correlators between chromo-electric and chromo-magnetic fields. These correlators can be computed non-perturbatively with lattice QCD (LQCD) \cite{Bali:1996bw,Bali:1996cj,Bali:1997am,Koma:2006fw,Koma:2007jq,Koma:2008zza,Koma:2009ws}.
    
    pNRQCD allows to determine the heavy quark-antiquark potential as an expansion in powers of $1/m$, where $m$ denotes the heavy quark mass~\cite{Brambilla:1999xf,Brambilla:1999xj,Brambilla:2000gk,Pineda:2000sz}. This potential can for example be used in the Schr\"odinger equation to predict masses of bottomonium and charmonium bound states (see e.g.\ Ref.\ \cite{Koma:2012bc}, where terms proportional to $1/m$ and $1/m^2$ were taken into account).
    A related, but technically more complicated problem is to study quarkonium resonances and tetraquark candidates above the lowest meson-meson threshold. In addition to a quarkonium channel one has to consider one or more meson-meson channels. The corresponding coupled channel Schr\"odinger equation is significantly more difficult to derive and to solve.

    In recent work \cite{Bicudo:2019ymo,Bicudo:2020qhp,Bicudo:2022ihz} we studied $I = 0$ bottomonium in such a way using LQCD static potentials computed in the context of string breaking \cite{Bali:2005fu}. $1/m$ and $1/m^2$ corrections, however, were not considered, but it is our plan to include them in the future. For the quarkonium channel these corrections are expected to be the same as for ordinary quarkonium bound states discussed in the previous paragraph.
    The aim of this work is to prepare and test the setup to compute masses of quarkonium bound states using $1/m$ and $1/m^2$ corrections to the static potential (similar to what was done in Ref.\ \cite{Koma:2012bc}). Moreover, we explore the gain in accuracy due to these corrections by comparing to the comprehensive set of experimental results available below the lowest meson-meson threshold. This should provide important insights and constitute a first step to include such corrections also in studies of bottomonium and charmonium resonances and tetraquark candidates above the lowest meson-meson threshold.

\section{Heavy quark-antiquark potential}

    The potential of a heavy quark and a heavy antiquark of the same mass $m$ up to order $\mathcal{O}(1/m^2)$ is of the form
\begin{align}
    V(r) = & V^{(0)}(r) + \frac{1}{m} V^{(1)}(r) + \frac{1}{m^2} V^{(2)}(r, \mathbf{p}, \mathbf{L}, \mathbf{S}) + \mathcal{O}(1/m^3) \label{eq:total_pot} .
\end{align}
    $V^{(2)}$ contains the momentum operator $\mathbf{p}$, the orbital angular momentum operator $\mathbf{L}$ and the spin operator $\mathbf{S}$ and can be expressed in terms of eight scalar functions, including $V_1'(r)$ and $V_2'(r)$ discussed below (see Refs.\ \cite{Barchielli:1986zs,Pineda:2000sz} for details). $V^{(0)}$, $V^{(1)}$ and the eight scalar functions defining $V^{(2)}$ can be computed with LQCD by evaluating Wilson loop-like correlation functions with chromo-electric and chromo-magnetic field insertions. Corresponding results are presented in Refs.\ \cite{Koma:2006fw,Koma:2007jq}.
    
    For quark-antiquark separations $r \gtapprox 0.2 \, \text{fm}$ LQCD results for the static potential $V^{(0)}$ can be parameterized by the Cornell ansatz
\begin{align}
    \label{EQN001} V^{(0)}(r) = & -\frac{\alpha}{r} + \sigma r + C
\end{align}
    with parameters $\alpha > 0$, the string tension $\sigma$ and a constant shift $C$ containing the lattice spacing dependent self energy (see e.g.\ Ref.\ \cite{Karbstein:2018mzo}).
    $V^{(1)}$ and the functions $V_1'$, $V_2'$, ... appearing in $V^{(2)}$ are known perturbatively up to $\mathcal{O}(\alpha_s^3)$ and $\mathcal{O}(\alpha_s^2)$, respectively, in the strong coupling \cite{Peset:2015vvi}.
    These perturbative expressions as well as the so-called Gromes~\cite{Gromes:1984ma} and BBMP~\cite{Barchielli:1986zs} relations suggest parameterizations of the LQCD results for $V^{(1)}$, $V_1'$, $V_2'$, ... depending only on two parameters, $\alpha$ and $\sigma$, which already appear in the parameterization of $V^{(0)}$, i.e.\ in Eq.\ (\ref{EQN001}).

    Refs.\ \cite{Bali:1997am,Koma:2006fw} confirm that these parameterizations provide a rather accurate description of non-perturbative LQCD results. Exceptions are the two spin potentials $V_1'$ and $V_2'$, where we generalize the parameterizations according to $V_1'(r) = -\epsilon_1 \alpha / r^2 - (1 - \epsilon_2) \sigma$ and $V_2'(r) = (1 - \epsilon_1) \alpha / r^2 + \epsilon_2 \sigma$, introducing two additional parameters $\epsilon_1$ and $\epsilon_2$. This is motivated by Refs.\ \cite{Bali:1996bw,Koma:2006fw}, where LQCD results for $V_1'$ indicate a constant behavior for large $r$, but suggest $V_1' \sim 1/r^2$ for small $r$.


\section{Schr\"odinger equation and quarkonium masses}

To compute masses of quarkonium bound states, we first write the Hamiltonian for the relative coordinate of the heavy quark-antiquark pair as
\begin{align}
H = H^{(0)} + \delta H
\end{align}
with
\begin{align}
H^{(0)} = \frac{p^2}{m} + V^{(0)}(r) \quad , \quad \delta H = -\frac{p^4}{4m^3} + \frac{1}{m} V^{(1)}(r) + \frac{1}{m^2} V^{(2)}(r, \mathbf{p}, \mathbf{L}, \mathbf{S})
\end{align}
(certain terms appearing in $V^{(1)}$ and $V^{(2)}$ can be incorporated in $H^{(0)}$ by a redefinition of the parameters $m$, $\alpha$ and $\sigma$; for details we refer to Ref.\ \cite{Bali:2000gf}). Note, that we also include the leading order relativistic kinetic correction $-p^4 / 4 m^3$ in $\delta H$.
Then we solve the Schr\"odinger equation $H^{(0)} \psi_{nl}^{(0)}(r,\theta,\varphi) = E_{nl}^{(0)} \psi_{nl}^{(0)}(r,\theta,\varphi)$ with $\psi_{nlm}^{(0)}(r,\theta,\varphi) = (u_{nl}(r) / r) Y_{lm}(\theta,\varphi)$, which simplifies to an ordinary differential equation in the radial coordinate $r$ and can be solved numerically using a standard Runge-Kutta shooting method.
Energy eigenvalues of the Schr\"odinger equation with the full Hamilton operator $H = H^{(0)} + \delta H$ are approximated in first order time independent perturbation theory as $E_{nl}^{(0)} + \langle \delta H \rangle$.

To generate numerical results we use $\alpha = 0.29$, $\sigma = 0.23 \, \text{GeV}^2$ (corresponding to $r_0 \approx 0.48 \, \text{fm}$) and $\epsilon_1 = \epsilon_2 = 0.2$. With this choice the parameterizations $V^{(0)}$, $V^{(1)}$, $V_1'$, $V_2'$, ... consistently describe the lattice data from Refs.~\cite{Koma:2006fw,Koma:2009ws}. The bottom quark mass $m_b$ and the charm quark mass $m_c$ were crudely tuned to round numbers\footnote{A more precise tuning of quark masses does not lead to significant changes in the results.}, such that the resulting quarkonium spectra are in fair agreement with experimental data~\cite{Workman:2022ynf}. We note that our chosen quark masses $m_b = 5000 \, \textrm{MeV}$ and $m_c = 1600 \, \text{MeV}$ are slightly below the masses of the corresponding heavy-light mesons $B$ and $D$ and rather close to values used in quark models \cite{Godfrey:1985xj}.
The constant shift $C$ in $V^{(0)}$ was set by identifying the spin averaged $1S$ mass, i.e.\ $(m_{\eta_b(1S)} + 3 m_{\Upsilon(1S)}) / 4$ for bottomonium and $(m_{\eta_c(1S)} + 3 m_{J/\psi(1S)}) / 4$ for charmonium, with $0$.

\begin{figure}
    \includegraphics[height=4.12cm, page=1]{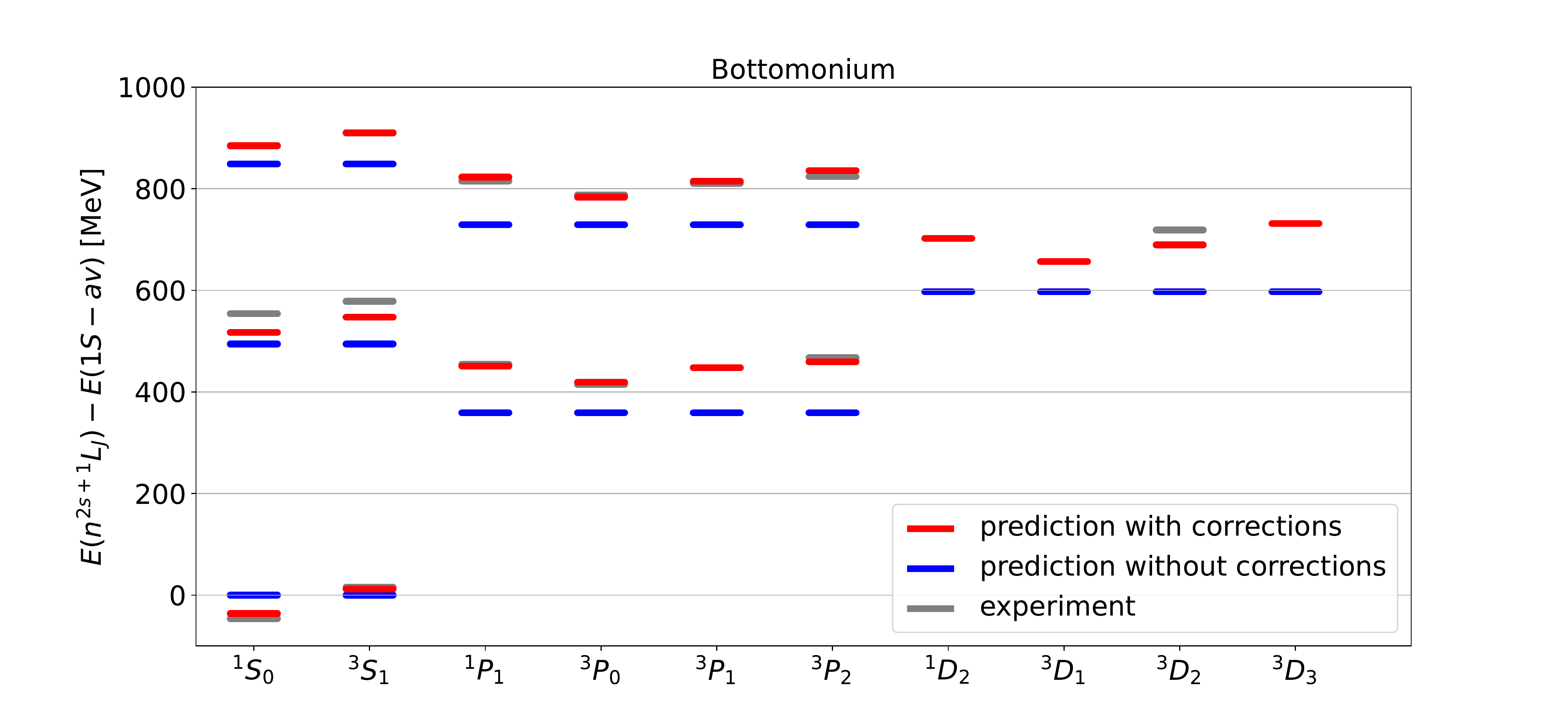}
    \includegraphics[height=4.12cm, page=1]{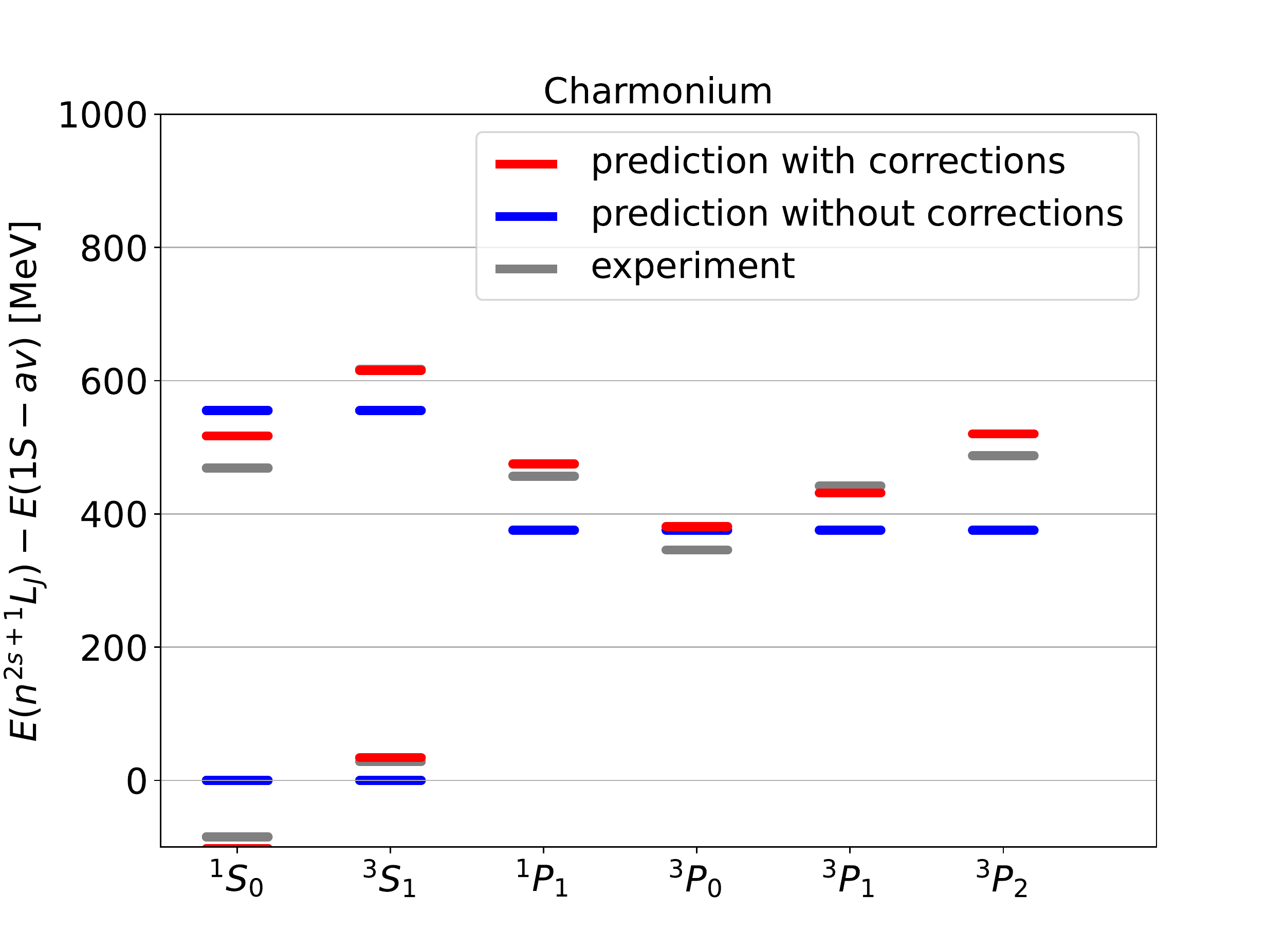}
        \caption{Bottomonium and charmonium masses computed with $1/m$ and $1/m^2$ corrections to the heavy quark-antiquark potential and the leading order relativistic kinetic correction (red lines), analogous results obtained without these corrections (blue lines) and corresponding experimental results (gray lines). States are labeled according to $^{2S+1}L_J$ ($J$ denotes total angular momentum).}
        \label{fig:bb_cc_spinensemble}
    \end{figure}

In Figure~\ref{fig:bb_cc_spinensemble} we compare our predicted bottomonium and charmonium masses (red lines) to their experimental counterparts (gray lines). We also show the energy levels $E_{nl}^{(0)}$, which represent crude estimates for bottomonium and charmonium masses based exclusively on the ordinary static potential $V^{(0)}$ (blue lines). The figure shows that $1/m$ and $1/m^2$ corrections to the heavy quark-antiquark potential and the leading order relativistic kinetic correction significantly improve the accuracy of our quarkonium predictions. The average difference between our theoretical results (blue and red lines) and existing experimental results (gray lines) is reduced by a factor of $\approx3$ for bottomonium and by a factor of $\approx4$ for charmonium.

In a next step we plan to treat the potential and relativistic kinetic corrections in a more stringent way, either by using higher order time independent perturbation theory or by deriving a coupled channel Schr\"odinger equation and projecting this equation to definite total angular momentum $J$. The latter would also constitute a first important step in the direction of studying $I = 0$ quarkonium resonances and tetraquarks, because it should be possible to combine such a coupled channel equation (coupled spin channels) with the coupled channel equation of Refs.\ \cite{Bicudo:2019ymo,Bicudo:2020qhp,Bicudo:2022ihz} (coupled quarkonium and meson-meson channels). This might allow to investigate not only bottomonium, but also charmonium resonances and tetraquarks above the lowest meson-meson threshold. Moreover, we are already in the process of carrying out an up-to-date precision LQCD computation of $V^{(0)}$, $V^{(1)}$, $V_1'$, $V_2'$, ... similar to what has been done in Refs.\ \cite{Bali:1996bw,Bali:1996cj,Bali:1997am,Koma:2006fw,Koma:2007jq,Koma:2008zza,Koma:2009ws} quite some time ago.

\section*{Acknowledgements}

    We acknowledge useful discussions with Pedro Bicudo, Nora Brambilla, Christian Fischer, Yanik Kleibrink and Lasse M\"uller. M.W.\ acknowledges support by the Heisenberg Programme of the Deutsche Forschungsgemeinschaft (DFG, German Research Foundation) -- project number 399217702.

\end{document}